\documentclass[12pt]{article}
\usepackage[english,german,french,polish]{babel}
\usepackage[T1]{fontenc}

\begin{document}

\baselineskip 0.75cm
\topmargin -0.4in
\oddsidemargin -0.1in

\let\ni=\noindent

\renewcommand{\thefootnote}{\fnsymbol{footnote}}

\newcommand{\CKM}{Cabibbo-Kobayashi-Maskawa }

\pagestyle {plain}

\setcounter{page}{1}

\pagestyle{empty}




~~~~~
\begin{flushright}
IFT-02/38
\end{flushright}

\vspace{0.2cm}

{\large\centerline{\bf  Does the LMA MSW solar solution imply}}

{\large\centerline{\bf  the Dirac nature of neutrinos?{\footnote {Work supported in part by the Polish State Committee for Scientific Research (KBN), grant 5 P03B 119 20 (2001--2002).}}}}

\vspace{0.5cm}

{\centerline {\sc Wojciech Kr\'{o}likowski}}

\vspace{0.23cm}

{\centerline {\it Institute of Theoretical Physics, Warsaw University}}

{\centerline {\it Ho\.{z}a 69,~~PL--00--681 Warszawa, ~Poland}}

\vspace{0.5cm}

{\centerline{\bf Abstract}}

\vspace{0.3cm}

Operating with the bilarge neutrino mixing, we show that in the option of Dirac neutrinos the fermion universality --- expressed by the proportionality of neutrino masses to charged-lepton masses --- predicts $\Delta m_{21}^2$  of the order $10^{-5}\;{\rm eV}^2$, consistently with the LMA MSW solar solution. In contrast, in the option of Majorana neutrinos the fermion universality --- introduced as the seesaw proportionality of neutrino masses to charged-lepton masses squared --- predicts $\Delta m_{21}^2$ of the order $10^{-8}\;{\rm eV}^2$, what is consistent rather with the LOW MSW solar solution. Thus, the favored LMA MSW estimation of $\Delta m_{21}^2$ might be a signal from the Dirac nature of neutrinos.

\vspace{0.4cm}

\ni PACS numbers: 12.15.Ff , 14.60.Pq , 12.15.Hh .

\vspace{0.8cm}

\ni October 2002

\vfill\eject

~~~~~
\pagestyle {plain}

\setcounter{page}{1}

\vspace{0.1cm}
 
As is well known [1], the bilarge form of mixing matrix for three active neutrinos $\nu_{e L}$, $\nu_{\mu L}$, $\nu_{\tau L}$ [1],

\vspace{-0.2cm}

\begin{equation} 
U =   \left( \begin{array}{ccc} c_{12} & s_{12} & 0 \\ -s_{12}c_{23} & c_{12}c_{23} & s_{23} \\ s_{12}s_{23} & -c_{12}s_{23} & c_{23} \end{array} \right) 
\end{equation}

\ni with $c_{ij} = \cos \theta_{ij}$ and $s_{ij} = \sin \theta_{ij}$, where $ \theta_{12} \sim 33^{\circ} $ (the favored LMA MSW solar solution) and $\theta_{23} \sim 45^{\circ}$, is globally consistent with the observed neutrino deficits for solar $\nu_e$'s [2] and atmospheric $\nu_\mu$'s [3] as well as with the negative Chooz experiment for reactor $\bar{\nu}_e$'s [4]. It cannot explain, however, the possible LSND effect for accelerator $\bar{\nu}_\mu$'s [5] that, if confirmed by the MiniBooNE experiment, may require the existence of a third neutrino mass-squared scale $\Delta m_{\rm LSND}^2 \sim1\;{\rm eV}^2$ beside the solar and atmospheric scales $\Delta m_{\rm sol}^2 \sim 5\times 10^{-5}\;{\rm eV}^2$ (the favored LMA MSW solar solution) and $\Delta m_{\rm atm}^2 \sim 3\times10^{-3}\;{\rm eV}^2$. The form (1) follows from the general shape {\it \`{a} la} \CKM of the neutrino mixing matrix [6] by putting $\theta_{13} = 0$, consistent with the negative result of Chooz experiment. The choice of $s_{13} = 0$ eliminates from $U$ the CP violating Dirac phase $\delta$ as it appears in $U$ only in the form $s_{13} \exp(\mp i\delta)$.

The neutrino mixing matrix $U = \left(U_{\alpha i}\right)$ defines the unitary transformation

\vspace{-0.2cm}

\begin{equation}
\nu_{\alpha L}  = \sum_i U_{\alpha i}  \nu_{i L} \;
\end{equation} 

\ni between the active-neutrino flavor and mass fields,  $\nu_{\alpha L}\;(\alpha = e, \mu, \tau)$ and $\nu_{i L}\;(i = 1,2,3)$, respectively. If the charged-lepton mass matrix is diagonal in the flavor representation, $U$ is at the same time the diagonalizing matrix for $M^\dagger M$, where $M = \left(M_{\alpha \beta}\right)$ is the neutrino mass matrix,

\vspace{-0.25cm}

\begin{equation} 
U^{\dagger} M^\dagger M U = {\rm diag}(|m_1|^2 \,,\,|m_2|^2  \,,\,|m_3|^2 )\,.
\end{equation} 

\ni Here $m_1 \,,\,m_2 \,,\,m_3$ denote generically complex neutrino masses, $m_1 = |m_1|$, $m_2 = |m_2|\exp(-2i\phi_2)$, $m_3 = |m_3|\exp(-2i\phi_3)$. We order $|m_1| \leq |m_2| \leq |m_3|$. 

There are two possible options: neutrinos are Majorana particles with the mass term $ -\frac{1}{2} \sum_{\alpha\,\beta}\overline{(\nu_{\alpha L})^c} M_{\alpha \beta} \nu_{\beta L}+{\rm h.c.}$, where $M$ is symmetric but generically complex, or they are Dirac particles with the mass term $ - \sum_{\alpha\,\beta} \overline{\nu_{\alpha R}} M_{\alpha \beta} \nu_{\beta L}+{\rm h.c.} =  - \sum_{\alpha\,\beta} \overline{(\nu_{\alpha L}+\nu_{\alpha R})} M_{\alpha \beta} (\nu_{\beta L}+\nu_{\beta R})$, where $M$ is Hermitian.

In the first option, Eq. (3) implies that the unitary $U$ diagonalizes symmetric $M$ in the complex orthogonal way

\vspace{-0.25cm}

\begin{equation} 
U^{T} M U = {\rm diag}(m_1 \,,\,m_2 \,,\,m_3)\,,
\end{equation} 

\ni where $ U^T = U^{\dagger *} = U^{* \dagger}$. Two phases $\phi_2$ and $\phi_3$ in $m_2$ and $m_3$ (thus present also in $M$) are known as Majorana phases. The complex orthogonal transformation inverse to Eq. (4) is

\vspace{-0.25cm}

\begin{equation} 
M =  U^* \, {\rm diag}(m_1\,,\,m_2\,,\,m_3)\, U^{* T} \,,
\end{equation} 

\ni {\it i.e.}, $M_{\alpha \beta} = \sum_i U^*_{\alpha i} m_i U^*_{\beta i}$. If the symmetric $M$ is real, then the unitary $U$ is real (as in the case of Eq. (1)), giving real neutrino masses $m_1 \,,\,m_2 \,,\,m_3$. Then, the CP violating Majorana phases are trivial. In general, these phases can be shifted in Eq. (4) from $m_1\,,\,m_2\,,\, m_3$ to $U$ replaced in such a way by $UP$ with $P = {\rm diag}(1\,,\,\exp(i\phi_2)\,,\,\exp(i\phi_3))$.

In the second option, Eq. (3) shows that the unitary $U$ diagonalizes Hermitian $M$ in the familiar unitary way

\vspace{-0.4cm}

\begin{equation} 
U^{\dagger} M U = {\rm diag}(m_1 \,,\,m_2  \,,\,m_3)\,,
\end{equation} 

\ni giving $m_1 \,,\,m_2 \,,\,m_3$ real (then, the phases $\phi_2$ and $\phi_3$ in $m_2$ and $m_3$ are trivial). The unitary transformation inverse to Eq. (6) is

\vspace{-0.25cm}

\begin{equation} 
M =  U \, {\rm diag}(m_1\,,\,m_2\,,\,m_3)\, U^\dagger 
\end{equation} 

\ni {\it i.e.}, $M_{\alpha \beta} = \sum_i U_{\alpha i} m_i U^*_{\beta i}$.

In the case of Eq. (1), where $U^* = U$, Eq. (5) or (7) leads (in both options) to the following mass matrix elements (with $m_1 \,,\,m_2 \,,\,m_3$ real, in the first option when $M^* = M$):

\vspace{-0.2cm}

\begin{eqnarray} 
M_{e e} & = & \;\,m_1 c^2_{12}+ m_2 s_{12}^2 \;, \nonumber \\
M_{\mu \mu} & = &\; (m_1 s^2_{12}+ m_2 c_{12}^2) c_{23}^2 + m_3 s^2_{23}\;, \nonumber \\
M_{\tau \tau} & = &\; (m_1 s^2_{12}+ m_2 c_{12}^2) s_{23}^2 + m_3 c^2_{23} \;, \nonumber \\ 
M_{e \mu} & = & \!\!\! -(m_1 - m_2) c_{12}s_{12} c_{23}\;\; = M_{\mu e}\;, \nonumber \\
M_{e \tau} & = & \; (m_1 - m_2) c_{12}s_{12} s_{23}\;\; = M_{\tau e}\;, \nonumber \\ 
M_{\mu \tau} & = & \!\!\!- (m_1s^2_{12}+ m_2c_{12}^2  - m_3) c_{23}s_{23} = M_{\tau \mu}\;,
\end{eqnarray} 

\ni where $c_{23} \sim 1/\sqrt{2} \sim s_{23}$ and $c_{12} \sim 0.84$ , $s_{12} \sim 0.54$.

For the bilarge form (1) of $U$ the following neutrino oscillation probabilities hold (in the vacuum):

\begin{eqnarray} 
P(\nu_e \rightarrow \nu_e) \; & = & 1 - (2c_{12}s_{12})^2 \sin^2 x _{21}\;, \nonumber \\
P(\nu_\mu \rightarrow \nu_\mu) & = & 1 - (2c_{12}s_{12})^2c^4_{23}\sin^2 x _{21} - (2c_{23}s_{23})^2 \left( s^2_{12}\sin^2 x _{31} + c^2_{12}\sin^2 x _{32}\right)\;, \nonumber \\
P(\nu_\mu \rightarrow \nu_e) & = & (2c_{12}s_{12})^2 c^2_{23} \sin^2 x _{21} \;, \nonumber \\
P(\nu_\tau \rightarrow \nu_e) & = & (2c_{12}s_{12})^2 s^2_{23} \sin^2 x _{21} \;, \nonumber \\
P(\nu_\mu \rightarrow \nu_\tau) & = & -(2c_{12}s_{12})^2 c_{23}^2 s_{23}^2 \sin^2 x _{21} + (2c_{23}s_{23})^2 (s_{12}^2 \sin^2 x _{31} + c_{12}^2 \sin^2 x _{32}) \;, 
\end{eqnarray} 

\ni where 

\begin{equation} 
x_{ji} \equiv 1.27 \frac{\Delta m^2_{ji} L}{E} \;,\; \Delta m^2_{ji} \equiv m^2_j - m^2_i \;\;(i,j = 1,2,3)
\end{equation} 

\ni ($\Delta m^2_{ji}$, $L$ and $E$ are measured in eV$^2$, km and GeV, respectively).  Here, $U^* = U$ and so, the possible CP violation is ignored. Thus, $P(\nu_\alpha \rightarrow \nu_\beta) = P(\bar{\nu}_\alpha \rightarrow \bar{\nu}_\beta) = P(\nu_\beta \rightarrow \nu_\alpha )$, the second equality following from the CPT theorem. In the formulae (9), $\Delta m^2_{21}\ll \Delta m^2_{32} \simeq \Delta m^2_{31}$ (and hence, $ x_{21} \ll x_{32} \simeq x_{31})$, since $\Delta m^2_{\rm sol} = \Delta m^2_{21}$ from the first Eq. (9) and $\Delta m^2_{\rm sol} \ll \Delta m^2_{\rm atm}$ experimentally, implying $\Delta m^2_{\rm atm} = \Delta m^2_{32}$ from the second Eq. (9), where the term with $\sin^2 x_{21} $ can be ignored. Therefore, the hierarchy of $ m^2_1 \leq  m^2_2 \leq  m^2_3$ is "$\,\!$normal": $\Delta m^2_{21}\ll \Delta m^2_{32} \simeq \Delta m^2_{31}$ (and not "inverse": $\Delta m^2_{31}\simeq \Delta m^2_{21} \gg \Delta m^2_{32}$).

In the option of Majorana neutrinos, postulating that the seesaw mechanism [7] works, we can write the mass matrix for active neutrinos $\nu_{e L}\,,\, \nu_{\mu L}\,,\, \nu_{\tau L}$ in the form

\begin{equation} 
M = -M^{(D)} \frac{1}{M^{(R)}}M^{(D)T} = U\, {\rm diag}(m_1\,,\,m_2\,,\,m_3)\, U^\dagger
\end{equation} 

\ni with the bilarge mixing matrix $U$ as given in Eq. (1). Here,

\begin{equation} 
{\cal L}_{\rm mass}^{(\nu)} = -\frac{1}{2}  \left( \overline{(\nu_{L})^c} \,,\, \overline{\nu_{R}}\right) \left( \begin{array}{cc} 0 & M^{(D)} \\ M^{(D)\,T} & M^{(R)} \end{array} \right) \left( \begin{array}{c} \nu_{L} \\ (\nu_{R})^c \end{array} \right) + {\rm h.\,c.} 
\end{equation} 

\ni with $\nu_{L} = \left( \nu_{e L}\,,\,\nu_{\mu L}\,,\,\nu_{\tau L} \right)^T$ and $ (\nu_{R})^c = \left( (\nu_{e R})^c\,,\, (\nu_{\mu R})^c\,,\, (\nu_{\tau R})^c \right)^T$ is the generic mass term for Majorana neutrinos, where $\nu_{\alpha L}$ and $(\nu_{\alpha R})^c \;(\alpha = e\,,\,\mu\,,\,\tau)$ are the active and (conventional) sterile neutrinos, respectively. In the simplest effective seesaw model, we put  [8] 

\begin{equation} 
M^{(D)} = U\;{\rm diag}(\lambda_1,\lambda_2,\lambda_3) U^\dagger \,,\, M^{(R)} = -U \,{\rm diag}( \Lambda, \Lambda, \Lambda) U^\dagger= -\Lambda {\bf 1}
\end{equation} 
 
\ni with $0 < \lambda_i \ll \Lambda $ (here, in the case of large mass scale $\Lambda$ of $M^{(R)}$, possible differences between its three eigenvalues are assumed to be negligible). Then, from Eq. (11)

\begin{equation} 
m_i =  \frac{\lambda^2_i}{\Lambda}\;\;\;(i = 1,2,3)\;.
\end{equation} 

Introducing in this option the fermion universality for neutrinos and charged leptons by the straightforward assumption that in Eq. (14) $\lambda_i $ are charged-lepton masses,

\begin{eqnarray}
\lambda_1 & = & m_e = 0.510999\,\;{\rm MeV}\;, \nonumber \\
\lambda_2 & = & m_\mu = 105.658\,\;{\rm MeV}\;, \nonumber \\
\lambda_3 & = & m_\tau = 1777.03\,\;{\rm MeV}\;, 
\end{eqnarray}

\ni we  obtain from Eq. (14)

\begin{equation}  
m_1 = 2.611\times 10^{-1}\, \frac{{\rm MeV}^2}{\Lambda} \;,\; m_2 = 1.116\times 10^4\, \frac{{\rm MeV}^2}{\Lambda} \;,\;m_3 = 3.158\times 10^6\, \frac{{\rm MeV}^2}{\Lambda}
\end {equation}  

\ni and


\begin{eqnarray}  
\Delta m^2_{21} = m^2_2 - m^2_1= 1.246\times 10^8 \frac{{\rm MeV}^4}{\Lambda^2}\!\! & , &\!\!  \Delta m^2_{32} = m^2_3 - m^2_2  = 9.972 \times 10^{12} \frac{{\rm MeV}^4}{\Lambda^2} \;, \nonumber \\
\Delta m^2_{21} /\Delta m^2_{32}\!\!  & = &\!\!  1.250\times 10^{-5}\;.
\end{eqnarray}   

\ni Making use of the SuperKamiokande estimate $\Delta m^2_{32}  \sim 3\times 10^{-3}\;{\rm eV}^2$, we {\it predict} from Eqs. (17) that 

\begin{equation}  
\Lambda \sim 5.8 \times 10^{10} \;{\rm GeV}\;,\; \Delta m^2_{21} \sim 3.7\times 10^{-8} \;{\rm eV}^2\;.
\end {equation}  

\ni The prediction (18) for $\Delta m^2_{21}$ [8] lies not very far from the experimental estimate $\Delta m^2_{21} \sim 7\times 10^{-8}\;{\rm eV}^2$ based on the LOW MSW solar solution, whereas the favored experimental estimation based on the LMA MSW solar solution is much larger: $\Delta m^2_{21} \sim 5\times 10^{-5}\;{\rm eV}^2$. The estimate (18) for $\Lambda$ gives from Eqs. (16) the neutrino masses

\begin{equation}  
m_1 \sim 4.5\times 10^{-9}\;{\rm eV}\,,\, m_2 \sim 1.9\times 10^{-4}\;{\rm eV}\,,\, m_3 \sim 5.5\times 10^{-2}\;{\rm eV}\,.
\end {equation}

In the option of Dirac neutrinos we can write the neutrino mass matrix in the form 

\begin{equation} 
M =M^{(D)} = U\, {\rm diag}(m_1 \,,\,m_2 \,,\,m_3)\,U^\dagger
\end{equation} 

\ni with the bilarge mixing matrix $U$ as presented in Eq. (1). 

Expressing in this option the fermion universality for neutrinos and charged leptons by the straightforward assumption of proportionality

\begin{equation}  
m_i =  \zeta \lambda_i \;\;(i = 1,2,3) ,
\end {equation}  

\ni where $\lambda_i$ are charged-lepton masses as given in Eqs. (15), we get

\begin{eqnarray}  
\Delta m^2_{21} = m^2_2 - m^2_1= 1.116\times 10^4 \zeta^2 \,{\rm MeV}^2 \!\! & , &\!\!  \Delta m^2_{32} = m^2_3 - m^2_2  = 3.147 \times 10^6 \zeta^2 \,{\rm MeV}^2 \;, \nonumber \\
\Delta m^2_{21} /\Delta m^2_{32}\!\!  & = &\!\!  3.548\times 10^{-3}\;.
\end{eqnarray}

\ni With the use of SuperKamiokande estimate $\Delta m^2_{32} \sim 3\times 10^{-3}\;{\rm eV}^2$ we {\it predict} from Eqs. (22) that 

\begin{equation}  
\zeta \sim 3.1 \times 10^{-11} \;,\; \Delta m^2_{21} \sim 1.1\times 10^{-5} \;{\rm eV}^2\;.
\end {equation}  

\ni The prediction (23) for $\Delta m^2_{21}$ is not very different from the favored experimental estimation $\Delta m^2_{21} \sim 5\times 10^{-5}\; {\rm eV}^2$ based on the LMA MSW solar solution. The estimate (23) for $\zeta $ implies by means of Eqs. (21) the neutrino masses

\begin{equation}  
m_1 \sim 1.6\times 10^{-5}\;{\rm eV}\,,\, m_2 \sim 3.3\times 10^{-3}\;{\rm eV}\,,\, m_3 \sim 5.5\times 10^{-2}\;{\rm eV}\,.
\end {equation}

In conclusion, the straigthforward formulation (21) of the fermion universality for neutrinos and charged leptons implies in the option of Dirac neutrinos the {\it prediction} of $\Delta m^2_{21}$ of the order $10^{-5}\;{\rm eV}^2$, not inconsistent with its most probable LMA MSW estimation $5\times10^{-5}\;{\rm eV}^2$. In contrast, the straigthforward seesaw formulation (14) and (15) of this universality leads in the option of Majorana neutrinos to the {\it prediction} of $\Delta m^2_{21}$ of the order $10^{-8}\;{\rm eV}^2$, not inconsistent with its less probable LOW MSW estimate $7\times 10^{-8}\;{\rm eV}^2$. Thus, on the ground of fermion universality, the favored LMA MSW estimation might be considered as a signal from the Dirac nature of neutrinos.

Notice finally that the charged-lepton mass spectrum (15) can be neatly parametrized by means of mass matrix of the form [9]:

\begin{equation}
M^{(e)} = \frac{1}{29} \left(\begin{array}{ccc} \mu^{(e)}\varepsilon^{(e)} & 0 & 0 
\\ 0 & 4\mu^{(e)}(80 + \varepsilon^{(e)})/9 & 0  
\\ 0 & 0 & 24\mu^{(e)} (624 + \varepsilon^{(e)})/25 \end{array}\right) 
\end{equation}

\ni which predicts $m_\tau = 1776.80$ MeV ({\it versus} $m^{\rm exp}_\tau = 1777.03^{+0.30}_{-0.26}$ MeV [10]), when the experimental values of $m_e$ and $m_\mu$ are used as an input (then, $\varepsilon^{(e)} = 0.172329$ and $\mu^{(e)} = 85.9924$ MeV are also determined). The fermion universality for neutrinos and charged leptons may be formulated in a straigthforward way by the assumption that the Dirac mass matrix $M^{(D)}$ for neutrinos gets --- after its diagonalization $U^\dagger M^{(D)} U $ is performed --- the form (25) with the parameters  $\mu^{(e)}$ and $\varepsilon^{(e)}$ replaced by new ones, $\mu^{(\nu)}$ and $\varepsilon^{(\nu)}$. This leads to the prediction of $\Delta m^2_{21}$ of the order $10^{-8}\;{\rm eV}^2$ or $10^{-5}\;{\rm eV}^2$, if neutrinos are Majorana or Dirac, respectively, and the estimate $\Delta m^2_{32} \sim 3\times10^{-3}\;{\rm eV}^2$ is used. Then, $\mu^{(\nu)\,2}/\Lambda \sim 1.3\times10^{-4}\;{\rm eV}$, or $\mu^{(\nu)} \sim 2.7\times10^{-3}\;{\rm eV}$, respectively. If in both options $\varepsilon^{(\nu)}$ and $\varepsilon^{(e)}$ are neglected, then $m_1 =0$ and $m_{2,3} = \left(\mu^{(\nu)}/\mu^{(e)}\right)^2 \lambda_{2,3}^2/ \Lambda$ with $\Lambda \sim 5.8\times 10^{10}$ GeV when $\mu^{(\nu)} = \mu^{(e)} $, or $m_1 = 0$ and $m_{2,3} = \zeta \lambda_{2,3}$ with $\zeta = \mu^{(\nu)}/\mu^{(e)} \sim 3.1\times 10^{-11}$, respectively. Here, $\lambda_i$ are charged-lepton masses as in Eq. (15). The form (25) of fermion mass matrix has a theoretical background described recently in Ref. [9], connected with the origin of three and only three fundamental fermion generations.

\vfill\eject

~~~~
\vspace{0.5cm}

{\centerline{\bf References}}

\vspace{0.45cm}

{\everypar={\hangindent=0.7truecm}
\parindent=0pt\frenchspacing

{\everypar={\hangindent=0.7truecm}
\parindent=0pt\frenchspacing

[1]~~For recent reviews {\it cf:} C. Giunti, {\it Talk at the 31st International Conference on High Energy Physics, July 2002, Amsterdam}, {\tt hep--ph/0209103}; M.C. Gonzalez-Garcia and Y.~Nir, {\tt hep--ph/0202056}.

\vspace{0.2cm}

[2]~~V. Barger, D. Marfatia, K. Whisnant and B.P.~~Wood, {\tt hep-ph/0204253}; J.N.~Bahcall,
M.C. Gonzalez--Garcia and C. Pe\~{n}a--Garay, {\tt hep--ph/0204314v2}; G.L.~Fogli {\it et al.}, {\tt hep--ph/0208026}; and references therein.

\vspace{0.2cm}

[3]~~S. Fukuda {\it et al.}, {\it Phys. Rev. Lett.} {\bf 85}, 3999 (2000).

\vspace{0.2cm}

[4]~~M. Appolonio {\it et al.}, {\it Phys. Lett.} {\bf B 420}, 397 (1998); {\bf B 466}, 415 (1999).

\vspace{0.2cm}

[5]~~G. Mills, {\it Nucl. Phys. Proc. Suppl.} {\bf 91}, 198 (2001); {\it cf.} also {\it cf.} K.~Eitel,  {\it Nucl. Phys. Proc. Suppl.} {\bf 91}, 191 (2001) for negative results of KARMEN2 experiment .

\vspace{0.2cm}

[6]~~Z. Maki, M. Nakagawa and S. Sakata, {\it Prog. Theor. Phys.} {\bf 28}, 870 (1962).

\vspace{0.2cm}

[7]~~M. Gell-Mann, P. Ramond and R.~Slansky, in {\it Supergravity}, edited by F.~van Nieuwenhuizen and D.~Freedman, North Holland, 1979; T.~Yanagida, Proc. of the {\it Workshop on Unified Theory and the Baryon Number in the Universe}, KEK, Japan, 1979; R.N.~Mohapatra and G.~Senjanovi\'{c}, {\it Phys. Rev. Lett.} {\bf 44}, 912 (1980).

\vspace{0.2cm}

[8]~~W. Kr\'{o}likowski, {\tt hep-ph/0208210}.

\vspace{0.2cm}

[9]~~W. Kr\'{o}likowski, {\it Acta Phys. Pol.} {\bf B 33}, 2559 (2002) (also {\tt hep--ph/0203107}); and references therein.

\vspace{0.2cm}

[10]~The Particle Data Group, {\it Eur. Phys. J.} {\bf C 15}, 1 (2000).

\vfill\eject

\end{document}